# Dendritic flux patterns in MgB$_2$ films

**T.H. Johansen\*, M. Baziljevich\*, D.V. Shantsev\*, P.E. Goa\*, Y.M. Galperin\*, W.N. Kang$^§$, H.J. Kim$^§$, E.M. Choi$^§$, M.-S. Kim$^§$, S.I. Lee$^§$**

*\*Department of Physics, University of Oslo, POB 1048 Blindern, N-0316 Oslo, Norway.*
*$^§$ National Creative Research Initiative Center for Superconductivity and Department of Physics, Pohang University of Science and Technology, Pohang 790-784, Republic of Korea.*

Correspondence: T.H. Johansen, e-mail: t.h.johansen@fys.uio.no

Magneto-opitcal studies of a *c*-oriented epitaxial MgB$_2$ film with critical current density $10^7$ A/cm² demonstrate a breakdown of the critical state at temperatures below 10 K. Instead of conventional uniform and gradual flux penetration in an applied magnetic field, we observe an abrupt invasion of complex dendritic structures. When the applied field subsequently decreases, similar dendritic structures of the return flux penetrate the film. The static and dynamic properties of the dendrites are discussed.

Following the recent discovery of superconductivity in polycrystalline MgB$_2$ [1], a tremendous effort is now being invested to produce high-quality films. Here we present studies of thin MgB$_2$ films with a record-high critical current density, $10^7$ A/cm² at 8 K. The films were deposited on ($1\bar{1}02$) Al$_2$O$_3$ substrates using pulsed laser deposition. Typical 400 nm thick films had a sharp superconducting transition ($\Delta T_C \sim 0.7$ K) at $T_C = 39$ K, and a high degree of c-axis alignment perpendicular to the plane [2]. Initial measurements of the integral magnetization vs. field displayed pronounced fluctuations at low $T$ indicating unusual features in the flux dynamics with onset between 20 and 5 K [3]. Spatially-resolved magneto-optical (MO) studies revealed [4] a complex dynamical behaviour of flux, where below $T = 10$ K the global penetration of vortices is dominated by dendritic structures abruptly entering the film. A similar dendritic behaviour was earlier observed in Nb-films [5], as well as in YBaCuO films when induced by a laser pulse [6]. In the present paper we investigate in more detail the properties of dendrites in films of $MgB_2$.

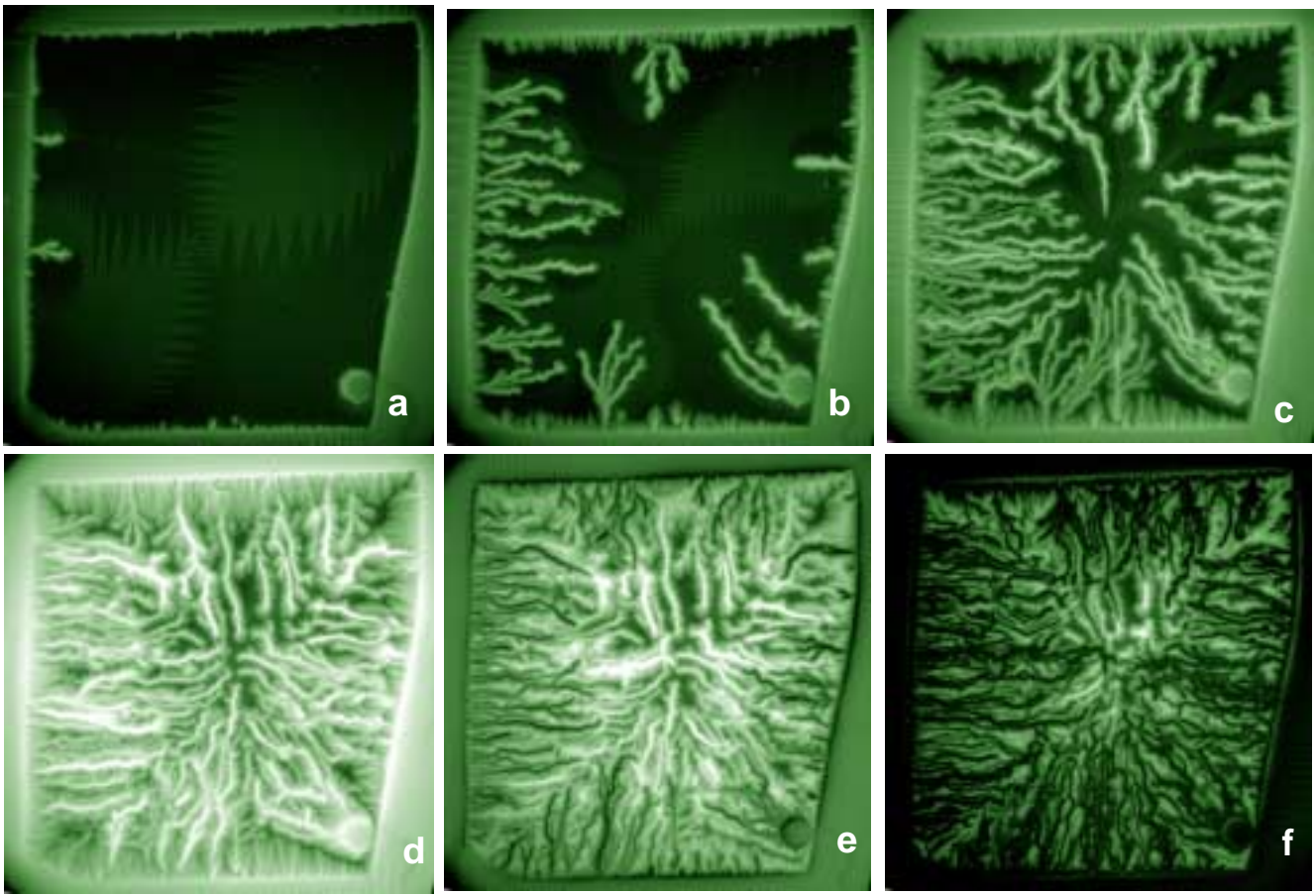

Figure 1. Magneto-optical images of flux penetration (image brightness represents flux density) into a zero-field-cooled MgB$_2$ film at 3 K. The respective images were taken at applied fields (perpendicular to the film) of 5, 10, 18, 35, 18, and 0 mT [7].

Figure 1 shows a sequence MO images taken as the applied field increases from 0 to 35 mT after first zero-field-cooling the sample to 3 K. The sweeping rate of the applied field was slow, of the order of 0.3 mT/sec, however the growth of the dendritic structures proceeded extremely fast: each structure was formed during less than 1 ms (the time resolution of our CCD system). The dendrite pattern never repeated itself in detail during a new experimental run under the same conditions.

The brightness on MO images represents the absolute value of the flux density. The dendrites formed at a larger applied field are seen to be brighter i.e. contain more flux than the ones formed at a lower field. However, once formed, each dendrite appears frozen and does not gain more flux as the applied field increases and other dendrites appear. Remarkably, the flux density in the dendrite core can sometimes be larger than the flux density at the film edge. This can be understood as a manifestation of strong demagnetization effects like in a case of an enhanced flux density at a defect near the film edge.

New dendrites while growing tend to avoid crossing the existing ones as illustrated by Fig. 2. The arrows in image (a) show where the new dendrite changed the growth direction because of "repulsion" from others that had appeared there before. Several strong turns were necessary to struggle through to the center of the film. Image (b) is obtained by subtraction of two MO images before and after the dendrite appeared. Dark regions here indicate a decrease of flux density. One can thus see that the new dendrite has consumed some vortices from the existing dendrites at places where passed by at close distance.

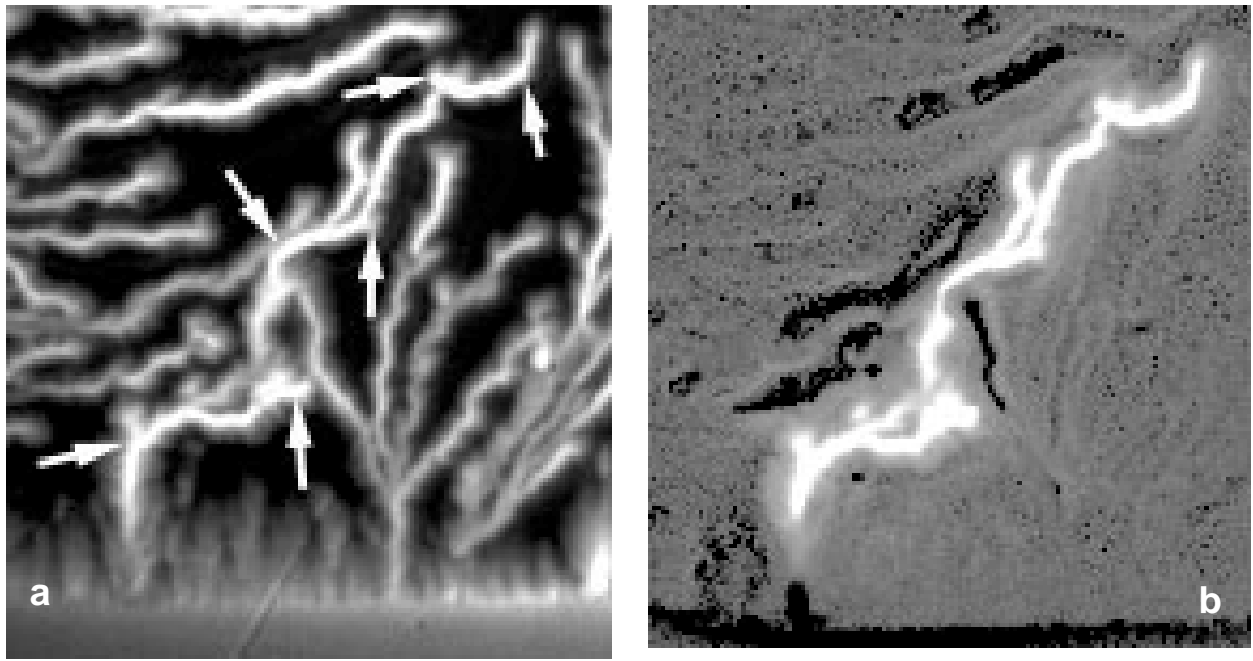

Figure 2. (a) Magneto-optical image of a region of MgB2 film. The dendrite that appeared last had to turn the growth direction several times (indicated by arrows) to avoid crossing the existing dendrites. (b) Result of subtraction of image (a) and MO image before appearing the dendrite. The grown dendrite is seen white, while the black regions indicate branches of existing dendrites affected by appearing the new one.

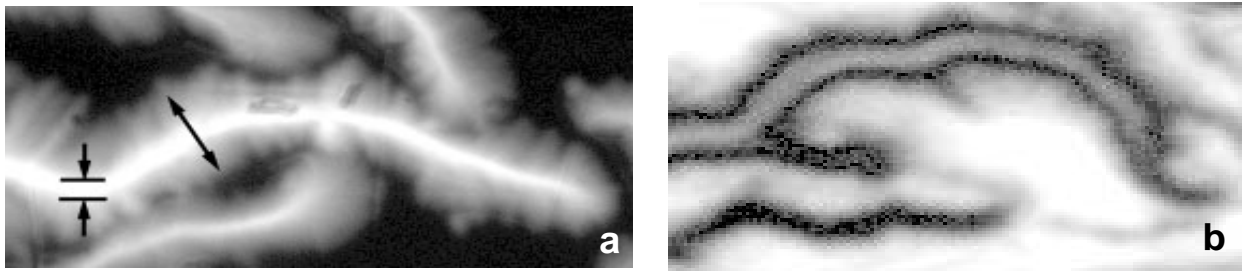

Figure 3. Blow-up of MO image showing individual dendrites. (a) A dendrite grown during increasing applied field. It consists of core, 20 μm wide, with a maximal flux density, and walls with flux density gradient summing up to the total dendrite width of 100 μm. (b) A dendrite grown during decreasing applied field. Its core contains flux of the opposite polarity which is separated from the surrounding flux by the annihilation zone seen black.

When the applied field is decreased from the maximal value back to zero, the flux redistributes in the same abrupt manners, see Fig.1 (e,f). Now, dendrites containing anti-flux of the reverse return field created by the trapped vortices penetrate the film. As a result, the remanent state, Fig.1(f), consists of an overlapping mixture of two types of dendrites of the opposite polarity. A closer look at the microstructure of the dendrites is given by Fig. 3. The dendrite of the initial polarity (a) consists of a core, ~20-30μm, with high flux density, and walls with a large field gradient. The total width of a dendrite can be 100 μm. Time-resolved studies of a similar phenomenon in YBaCuO films show that at first the core of the dendrite grows to its full length, and then the walls are formed where a sort of conventional critical-state flux profile is established. It is worthwhile to note however, that several dendrites have a different microstructure: the middle of the core had a lower flux density [4]. A dendrite shown in image (b) was formed during the field decrease and contains anti-flux in its core. The black annihilation zone separating the anti-flux from the surrounding flux of initial polarity is clearly seen.

We observed the dendritic flux penetration into the virgin state of $MgB_2$ film only for temperatures below 10 K. At higher temperatures the penetration proceeded in a conventional smooth and uniform way. However, after the applied field was decreased from the maximal value, dendrites could be seen even in the temperature range between 10 and 13 K. This indicates that the decreasing-field state is less stable with respect to dendrite formation. A possible reason for this is an existence of an annihilation zone, where extra heat is released due to annihilation of vortices and anti-vortices. Peculiar instabilities near annihilation zone have been observed earlier in HTSC single-crystals [8]. This observation supports the conclusion of Ref. [4] that the dendritic flux penetration originated from a thermo-magnetic instability.